\documentclass[a4paper,UKenglish]{lipics}
 
\usepackage{microtype}

 \def\map#1{\mathcal #1}
\def\d{\operatorname{d}}\def\<{\langle}\def\>{\rangle}
\def\Tr{\operatorname{Tr}}\def\:{\hbox{\bf:}}
\def\Cmplx{\mathbb C}
\def\spc#1{\mathscr{#1}}

\def\grp#1{\mathsf{#1}}
\def\Lin{\mathsf{Lin}}
\title{Is global asymptotic cloning state estimation?
}
\author[1]{Yuxiang Yang}
\author[2]{Giulio Chiribella}
\affil[1]{Department of Physics, Tsinghua University\\
  Beijing, 100084, China\\
  \texttt{yangyx09@mails.tsinghua.edu.cn}}
\affil[2]{Center for Quantum Information, Institute for Interdisciplinary Information Sciences, Tsinghua University\\
  Beijing, 100084, China\\
  \texttt{gchiribella@mail.tsinghua.edu.cn}}
\authorrunning{Yuxiang Yang and Giulio Chiribella} 

\Copyright[by]{Yuxiang Yang and Giulio Chiribella}


\serieslogo{}
\volumeinfo
  {Billy Editor, Bill Editors}
  {2}
  {Conference title on which this volume is based on}
  {1}
  {1}
  {1}
\EventShortName{}
\DOI{10.4230/LIPIcs.xxx.yyy.p}

\begin{document}
\maketitle
\begin{abstract}
We investigate the asymptotic relationship between quantum cloning and quantum estimation from the global point of view where all the copies produced by the cloner are considered jointly. 
For an $N$-to-$M$ cloner, we consider the overall  fidelity between the state of the $M$ output systems and the state of $M$ ideal copies, and we ask whether the optimal fidelity  is attained by a  measure-and-prepare protocol in the limit $M\to\infty$.    In order to gain intuition into the general problem, we analyze  two  concrete examples: \emph{i)} cloning qubit states on the equator of the Bloch sphere and \emph{ii)} cloning two-qubit maximally entangled states.   In  the first case, we show that  the optimal measure-and-prepare fidelity converges to the fidelity of the optimal cloner in the limit $M\to \infty$.   In the second case, we restrict our attention to economical covariant cloners, and again, we exhibit a measure-and-prepare protocol that achieves asymptotically the optimal fidelity.             
Quite counterintuitively, in both cases the optimal states that have to be prepared in order to maximize the overall fidelity are not product states corresponding to $M$ identical copies, but instead suitable $M$-partite entangled states.     
 \end{abstract}

\section{Introduction}

It is well known that every quantum machine producing a large number of indistinguishable clones---referred to as \emph{asymptotic cloning machine}---is ``equivalent" to  a machine that measures the input states and re-prepares many identical copies of a state depending on the outcome  \cite{BrussEkert98,BaeAcin06,ChiribellaDariano06,Chiribella11}.   Here, ``equivalent" has to be understood  in the following  sense:  when one restricts the attention to a few clones, their state will be almost indistinguishable from the state that can be produced by a measure-and-prepare protocol.   Precisely, the trace distance between the state of  $k$ clones produced by machine and the state of $k$  clones produced by the  measure-and-prepare protocol goes to zero as $k/M$, where $M$ is the number of output copies \cite{ChiribellaDariano06,Chiribella11}.
For $k=1$, the fact that  the state of each  individual clone is asymptotically equal to the state produced by a measure-and-prepare protocol implies that the single-copy fidelity of quantum cloning is asymptotically equal to the fidelity of state estimation, a fact that is commonly known as ``equivalence between asymptotic cloning and state estimation" \cite{KeylWebsite}.   
 
In this paper we raise the question whether the equivalence between asymptotic cloning and state estimation continues to hold when one considers all the $M$ clones together, rather than restricting the attention to a single clone or a small subset of $k$ clones.   
We refer to this new form of  equivalence as \emph{global asymptotic equivalence between quantum cloning and state estimation} and we conjecture that the equivalence holds. 
 A few observations supporting the conjecture are the following:   
First of all, in the known cases---cloning of arbitrary pure states \cite{BrussDivincenzo98,Werner98} and the cloning  of coherent states \cite{CerfIpe00,Lindblad00,CerfIblisdir00,CochraneRalph04,ChiribellaXie13}---the equivalence holds, and in a fairly strong sense:  the joint state of all the  output clones  converges in trace distance to the output state of a measure-and-prepare protocol.   
A more general argument supporting our conjecture comes from the intuition that  producing a large number of identical copies means ``classicalizing" the information contained in the input states, and therefore it is natural to expect that the optimal way to classicalize quantum information is to perform a measurement.

In order to discuss the question of the global equivalence  one needs first to fix the rules of the game, by defining  a suitable figure of merit.   
Here we consider  the \emph{global fidelity}, namely the overlap between the output state of all clones  and  the desired quantum state of $M$ identical copies. 
In this setting, proving the equivalence means proving that the global fidelity of the optimal $N$-to-$M$ cloner  can be  achieved by a measure-and-prepare protocol in the asymptotic limit $M \to \infty$.   
In order to gain intuition into the problem, we  consider two concrete examples: the cloning of qubit states on the equator of the Bloch sphere and the cloning of two-qubit maximally entangled states.  In the first case it is known that the optimal cloner, derived in Ref. \cite{DarianoMacchiavello03}, is  \emph{economical} \cite{NiuGriffiths99,BuscemiDariano05,DurtFiurasek05}, that is, it can be implemented by a unitary interaction between the $N$ input copies and  $M-N$ blank copies.  An economical cloner is far from being implementable by a measure-and-prepare protocol, and observing an asymptotic equality of fidelities becomes here a quite non-trivial matter.   In the second case (cloning of maximally entangled states), we  will deliberately restrict ourselves to economical cloning machines,  asking the question whether the global fidelity of the optimal economical cloner can be achieved by measurement and re-preparation.   
In both cases we will give an affirmative answer, showing that the difference between the global fidelity of the optimal economical $N$-to-$M$ cloner and the global fidelity of the optimal measure-and-prepare protocol  becomes negligible in the asymptotic limit $M \to \infty$, for every fixed value of $N$. Quite counterintuitively, we observe  that the obvious protocols consisting in estimation of the unknown state and re-preparation of $M$ identical copies do not reach the maximum fidelity, \emph{even in the asymptotic limit.} 
This feature is in stark contrast with the intuition coming from the single-copy scenario, where re-preparing identical copies of the same state is asymptotically the best strategy. 

\section{Preliminaries}\label{sec:preliminary}

In this section we formalize the problem of the joint asymptotic equivalence and give an overview of the methods used in the rest of the paper. 
 \subsection{The problem of the global asymptotic equivalence}  
Consider a set of states $\{ |\psi_x\> \}_{x\in\mathsf X}$ in a finite dimensional Hilbert space $\spc H$.      
  The task of optimal quantum cloning  is to convert $N$ perfect copies of an unknown state $|\psi_x\>$,  given with probability $p_x$, into $M$ approximate copies that are as accurate as possible.  
  Examples of this problem are the universal cloning of pure states \cite{BuzekHillery96,
  BuzekBraunstein97,GisinMassar97,BrussDivincenzo98,Werner98,KeylWerner99} and the phase-covariant cloning  \cite{BrussCinchetti00,DarianoMacchiavello03,FanMatsumoto01,FanMatsumoto02,BuscemiDariano05,DurtFiurasek05}.
  
The most general  cloning process will be described by a quantum channel (completely positive trace-preserving map) $\map C$  transforming density matrices on $\spc H^{\otimes N}$  to density matrices on $\spc H^{\otimes M}$. As a figure of merit for the quality of the copies we will consider the \emph{global fidelity},  
\begin{equation}\label{generalfid}
F[N \to M]=\sum_{x\in\mathsf X}   \Tr\left[  \psi_x^{\otimes M}\map C(\psi_x^{\otimes N})\right]  \qquad \psi_x  :  = | \psi_x \>\< \psi_x| .
\end{equation}
When the set $  \{  |\psi_x\>\}_{x\in\mathsf X}$ is continuous, it is understood that the sum over the possible input states has to replaced with an integral with a suitable probability distribution $p(x) ~dx$.     The optimal cloner will be the quantum channel that maximizes $F[N\to M]$.  The fidelity of the optimal cloner will be denoted by by $F_{clon}[N\to M]$.  

In addition to the maximum over all channels, it is important to  consider the maximum of $F[N\to M]$ over the set of \emph{measure-and-prepare channels}.  Operationally, a measure-and-prepare channel can be realized by measuring the input copies with a POVM $  ( P_y)_{y\in\mathsf Y}$  and, when the measurement gives outcome $y$, by re-preparing a state $\rho_y$.
  Averaging over the measurement outcomes, the action of the measure-and-prepare channel on the density matrices is given by $ \map C (\rho)  = \sum_{y\in\mathsf Y}     \Tr [   P_y \rho]  ~   \rho_y$.     We will denote by $F_{est} [N\to M]$ the maximum of the fidelity over the set of  measure-and-prepare channels.   Such a  maximum  is known in the literature as \emph{classical fidelity threshold} \cite{HammererWolf05,AdessoChiribella08,OwariPlenio08,CalsamigliaAspachs09,
ChiribellaXie13} and can be used as a benchmark for the experimental demonstration of quantum advantages. 
  
In the following we will ask the question whether the difference between $F_{clon}[N\to  M ] $ and $F_{est}[N\to M]$ becomes negligible in the asymptotic limit $M \to \infty$, while keeping $N$ fixed.   An affirmative answer to this question would mean that the quantum way to process information and the classical way fare equally well in the asymptotic limit.  
   In the formalization of the problem there is a catch, because both fidelities converge to zero in many interesting cases when the family of states to be cloned is continuous: a non-vanishing fidelity would indeed violate the Heisenberg limit of quantum metrology \cite{ChiribellaYang13}.  In order not to trivialize the question, it is then important to consider the \emph{relative} difference between the two fidelities, given by  
 \begin{align}
 \Delta [N\to M]  :  =  \frac{  F_{clon}[N \to M]   -  F_{est}[N  \to M]}{F_{clon}[N \to M]} .
  \end{align}   
 
Our conjecture is that, for every fixed $N$, the relative difference vanishes in the limit $M  \to \infty$. In formula: 
 \begin{align}\label{conj1}
 \lim_{M \to \infty} \frac{  F_{clon}[N \to M]   -  F_{est}[N  \to M]}{F_{clon}[N \to M]}=0 \qquad \forall N \in\mathbb N.
  \end{align}   
We refer to the conjectured equality as \emph{global asymptotic equivalence between quantum cloning and quantum state estimation}.    Of course, here the word ``global" refers to the fact that we are considering the global fidelity as the performance measur, as opposed to the single-copy fidelity considered in the previous literature.   \cite{GisinMassar97,BrussDivincenzo98,KeylWerner99} 

  From previous results on optimal cloning we know that the relation is satisfied in the case of universal quantum cloning \cite{BrussDivincenzo98,Werner98}  (see \cite{Chiribella11} for the proof that the optimal channel converges to a measure-and-prepare channel)  and in the case of the coherent-state quantum cloning \cite{CerfIpe00,Lindblad00,CerfIblisdir00,CochraneRalph04,ChiribellaXie13}   (see \cite{ChiribellaXie13} for the proof that  $F_{clon} [N \to M]$ becomes asymptotically equal to  $F_{est} [N\to M]$, up to a negligible error). In the following we will exhibit two new examples supporting the conjecture that joint cloning is asymptotically equivalent to state estimation. 
  
In the first example, we consider the optimal   cloning of qubit states on the equator of the Bloch sphere.  In this case, the optimal $N$-to-$M$ cloner is known  \cite{DarianoMacchiavello03}) and has a very interesting feature: it can be realized through a unitary interaction between the $N$ input copies and only $M-N$ blank copies of the input system.  In formula, the optimal quantum channel has the form  
\begin{align}
\map C  (\rho)  =    U   \left [\rho  \otimes |0\>\<0|^{\otimes (M-N)}\right]  U^\dag,
\end{align}  
where $U:  \spc H^{\otimes M} \to \spc H^{\otimes M}$ is a unitary operator and $|0\>$ is a fixed state in $\spc H$.   
Cloning channels of this form are usually referred to as \emph{economical} \cite{NiuGriffiths99,BuscemiDariano05,DurtFiurasek05}.  
  For the optimal cloner of  qubit states on the equator, we will show that our conjecture holds, by explicitly constructing a family of measure-and-prepare channels that attains the maximum  fidelity $F_{clon}[N\to M]$.       In a sense, this example is more intriguing than the previous ones, because the economical cloner considered here is far from being achieved by measure-and-prepare protocols: the asymptotic equivalence is then a non-trivial relation between the optimal joint fidelities.

In the second example, we consider the cloning of two-qubit maximally entangled states. For simplicity, here we restrict our attention to economical quantum cloners satisfying a natural symmetry requirement, and we denote by $F_{clon,eco} [N\to M]$ the maximum fidelity achieved by these channels.  In this case, we show that  the maximum value $F_{clon,eco}[N\to M]$ can be achieved by a suitable family of measure-and-prepare channels, in the limit $M\to N$.  Again, this example supports the validity of our conjecture.

\subsection{General methods}
Here we make some general considerations that apply to the two specific examples considered in   the paper.      

\subsubsection{Covariant economical channels} 
In many relevant cases,          the unknown state to be cloned is of the form  $|  \psi_g \>  :=  U_g   | \psi \>$ , where  $ |\psi\> \in \spc H$  is unit vector and  $U:  \grp G  \to \Lin (\spc H), g \mapsto U_g$ is a unitary representation of a compact group $\grp G$ on the set $\Lin (\spc H)$ of linear operators on $\spc H$.  
Examples of this problem are the universal cloning of pure states \cite{Werner98} and the phase-covariant cloning  \cite{BrussCinchetti00,DarianoMacchiavello03,FanMatsumoto01,FanMatsumoto02,BuscemiDariano05,DurtFiurasek05}. Due to the symmetry of the states, the maximum of the fidelity can be achieved by choosing a \emph{covariant  channel}, namely a quantum channel satisfying the property  
\begin{equation}\label{covariance}
\map C  \circ    \map U_{g}^{\otimes N}  =   \map U_{g}^{\otimes M}  \circ  \map C  \qquad g \in \grp G,
\end{equation} 
where $ \map U_{g}  $  is the unitary channel defined by $\map U_g (\rho)  =  U_g \rho  U_g^\dag$, for every density matrix $\rho$.   For covariant channels, the expression of the fidelity is reduced to  
\begin{equation}\label{covfid}
F[N \to M]=  \Tr \left[  \psi^{\otimes M}\map C(\psi^{\otimes N})\right]  .
\end{equation}

A further simplification arises if we assume that the covariant channel $\map C$ is economical, namely $\map C(\rho)  =  V \rho V^\dag$ for a suitable isometry $V$:  the fidelity takes the simple form 
\begin{align}\label{covfidiso}
F[N\to M]  = \left| \<  \psi|^{\otimes M}  V  |\psi\>^{\otimes N}  \right|^2 
\end{align}   
and  the covariance condition  becomes  
\begin{align}\label{covarianceiso}
U_g^{\otimes M}   V  \left( U_g^{\otimes N}\right)^\dag   =  \omega_g ~ V  \qquad g\in\grp G,
\end{align}
where $  \omega:   \grp G  \to \Cmplx$ is a one-dimensional representation of the group $\grp G$.  For the cloning of maximally entangled states of qudits, where the group is $SU(d)$, Eq. (\ref{covarianceiso}) is simply   
\begin{align}\label{covarianceiso2}
U_g^{\otimes M}   V  \left( U_g^{\otimes N}\right)^\dag   =   ~ V  \qquad g\in\grp G,
\end{align}
because the only one-dimensional representation of $SU(d)$ is the trivial one  ($  \omega_g  = 1   , ~ \forall g$).   
 
\subsubsection{Covariant measure-and-prepare protocols}\label{subsec:covmeasprep}

In order to prove the global asymptotic equivalence, our goal is to construct a family of measure-and-prepare protocols that attains the fidelity of the best quantum cloners in the  limit $M\to \infty$.    To achieve this goal, we will make a series of assumptions motivated by physical intuition. A posteriori, the fact that our protocols attain the desired fidelity will provide a confirmation that the intuition was sound. 


First of all, for an input state of the form $|\psi_g\>  =  U_g  |\psi\>$  we  will consider measure-and-prepare strategies that are based on state estimation, namely strategies where the set of measurement outcomes  coincides with the set parametrizing the input states, namely $\mathsf X \equiv  \grp G$.  
 Hence,  the measurement is described by a POVM  $  P_{\hat{g}}~  d\hat{g}$ with in the group $\grp G$, normalized as 
 $\int  \d \hat g  ~  P_{\hat g}  =  I^{\otimes N} $.

For the re-preparation stage , we will require that the  states that are re-prepared have the  form   $|\Phi_{\hat{g}} \>   = U^{\otimes M}_{\hat{g}} |\Phi \> $, for a given unit vector $|\Phi\>\in\spc H^{\otimes M}$.   
With this particular choice, the optimization of the measure-and-prepare  protocol is equivalent to the optimization of a state  estimation protocol that is designed to maximize the average of the  function 
\begin{align}
f( \hat g, g)   :  =  \Tr \left[ \Phi_{\hat{g}}  ~  \psi_g^{\otimes M} \right] .
\end{align} 
In this case, is known  that the optimal POVM can be chosen to be \emph{covariant} \cite{Holevo}, that is,  $P_{\hat{g}}=U_{\hat{g}}^{\otimes N}  \eta  U_{\hat{g}}^{\dag \otimes N}$ where $\eta   \in  \Lin \left(  \spc H^{\otimes N}\right)$ is a suitable positive operator, called the seed of POVM.  For a covariant POVM, the probability density $p(\hat g  |  g)  =  \Tr [P_{\hat g}   \psi_g^{\otimes N} ]$ satisfies the relation  
\begin{align}
p(h \hat g |  h g)  =  p(\hat g| g)  \qquad \forall h, \hat g, g\in\grp G. 
\end{align}  
Hence, the fidelity of the corresponding measure-and-prepare protocol becomes 
\begin{equation}\label{eq:m&p-fidelity}
F [  N \to M]=\int \d g ~    \Tr \left[  \eta~  \psi_g^{\otimes N}\right]  ~  \Tr \left[ \Phi~  \psi_g^{\otimes M}\right]  .  
\end{equation}

Finding the optimal measure-and-prepare protocol is then reduced to finding the optimal  operator $ \eta$  and the optimal state $|\Phi\>$.   
 To this purpose, in the two examples considered in this paper we will  make a   
 suitable ansatz on the form of the state $|\Phi\>$, which guarantees that  $\Tr \left [\eta ~ \psi_g^{\otimes N}\right]$,  as a function of $g$, varies slowly  with respect to $\Tr \left[\Phi   ~\psi_g^{\otimes M}\right]$, which is concentrated around its maximum at $g  = e$, the identity element of the group.  Under this ansatz, the fidelity can be approximated as
\begin{equation}\label{eq:m&p-approx}
\begin{split}
F[N \to M]&\approx\left\{\int \d g ~  \Tr \left [ \Phi ~  \psi_g^{\otimes M}  \right]    \right\}~  \Tr \left[  \eta ~  \psi^{\otimes N}\right] =  \<\Phi|     \rho^{(M)}_{aver}  |\Phi\>~  p^{(N)}_{true}.
\end{split}
\end{equation}
where  $\rho_{aver}^{(M)}:=\int dg  ~\psi_g^{\otimes M}$ is the average state of $M$ ideal copies and $p^{(N)}_{true}:= \Tr \left[ \eta~ \psi^{\otimes N}\right] $ is the probability density that the estimated value  $\hat{g}$ coincides with true values $g$.   

Thanks to Eq. (\ref{eq:m&p-approx}),  optimizing the measure-and-prepare protocol is reduced to two independent optimization problems: the maximization of the fidelity between the state $|\Phi\>$ and the average state  $\rho_{aver}^{(M)}$ (under the restriction that $|\Phi\> $ must be compatible  with the ansatz) and the maximization of the probability density $p_{true}^{(N)}$.   
In the specific cases considered in this paper, we will show that the ansatz can be done without loss of generality: indeed, the fidelity achieved by measure-and-prepare protocols satisfying the ansatz  approaches the fidelity  of the optimal quantum channel.  

\section{Cloning equatorial qubit states }\label{sec:qubit}
Here we  consider the optimal $N$-to-$M$ cloning of pure qubit states on the equator of the Bloch sphere, evaluating the asymptotic expression of the optimal quantum fidelity and  showing that it can be  achieved via a suitable measure-and-prepare protocol.    

\subsection{The performance of the optimal quantum cloner}

Consider the qubit states on the equator of  the Bloch sphere,   defined as
\begin{align*}
|\psi_\theta\> &=  \frac{|0\>+e^{i\theta}|1\>}{\sqrt{2}}    \qquad \theta \in [-\pi,\pi)\\
&=U_{\theta} |\psi\>    \\
  U_{\theta}  &: =\exp \left[\frac {i\theta(\sigma_z+I)}2  \right] ,  \quad   |\psi\> :  =   \frac{|0\>+|1\>}{\sqrt{2}}.  \\
 \end{align*}
 The state of the $N$ input copies can be represented as 
 \begin{align*}
 |\psi\>^{\otimes N}    =    \sum_{n=  -N/2}^{N/2}    \sqrt{b_{N, n}  }  | N/ 2  ,  n  \>   \qquad b_{N,n}  :  =   \frac 1 {2^N}  {N\choose N/2+n}
 \end{align*} 
 where    $\{|N/2,n\>~|~  n  =  -N/2  ,\dots, N/2\}$ are the Dicke states and $b_{N,n}$ is the binomial distribution.

 The optimal cloning channel was derived in Ref.  \cite{DarianoMacchiavello03}.  When $M-N$ is even, the optimal channel  is covariant with respect to the action of the phase shifts $U_\theta$ and economical, i.e. of the form $\map C (\rho)  =  V \rho V^\dag$ where $V : \spc H^{\otimes N} \to \spc H^{\otimes M}$ is an isometry(i.e. $V^\dag V  =  I$).   Specifically, the isometry of the  optimal cloner  is   $V  =  \sum_{n  =  -N/2}^{  N/2}     | M/2, n\>  \< N/2 , n| $ and produces the output state
 \begin{equation}
V |\psi\>^{\otimes N}   = \sum_{m=-N/2}^{N/2}\sqrt{b_{N,m}}|M/2,m\>.  
 \end{equation}    
 Inserting this  expression in  Eq. \eqref{covfidiso} one gets the maximum fidelity \cite{DarianoMacchiavello03}
\begin{equation}\label{eq:qubitclone}
F_{clon}[N\to M]=\left(\sum_{n=-N/2}^{N/2}\sqrt{b_{N,n}b_{M,n}}\right)^2.
\end{equation}
When $M$ is large compared to $N$, the fidelity becomes:
\begin{equation}\label{eq:qubit opt fidelity}
F_{clon}[N \to M ]\approx b_{M,0}\left(\sum_{n=-N/2}^{N/2}\sqrt{b_{N,n}}\right)^2  \qquad M  \gg N.
\end{equation}
In the next subsection we will construct a measure-and-prepare channel that achieves this asymptotic value for every $N\in\mathbb N$.

Note that the optimal quantum fidelity in Eq. (\ref{eq:qubitclone}) has a  simple expression when $N$ is large ($N\gg 1$).  In this case, the probability distributions $b_{N,n}$  and $b_{M,n}$ are well approximated by the  Gaussian distributions $g_N(x)=\sqrt{\frac{2}{\pi N}}e^{-2x^2/N}$ and  $g_M (x)=\sqrt{\frac{2}{\pi M}}e^{-2x^2/M}$, respectively.  Replacing  the summation in Eq. \eqref{eq:qubitclone} with a Gaussian integral,  one gets
\begin{equation}
F_{clon} [  N  \to M] \approx \frac{  \sqrt{  4 M N} }{M+N} \qquad    N \gg 1.
\end{equation} 
Incidentally, it is interesting to observe that in this regime the fidelity is close to 1 whenever  the number of extra-copies $M-N$ is negligible compared to $N$, whereas it is close to 0 whenever $N$ is negligible compared to $M$.  This fact is an illustration of the standard quantum limit for cloning introduced in Ref. \cite{ChiribellaYang13}.  

\subsection{A family of measure-and-prepare protocols achieving asymptotically the optimal fidelity}\label{subsec:qubitmeasprep}
Here we consider the maximization of the cloning fidelity over measure-and-prepare  protocols based on state estimation (cf.  subsection \ref{subsec:covmeasprep}). 
For equatorial qubit states, the measure-and-prepare protocol consists in the estimation of the parameter  $\theta  \in  [0,2\pi)$ from the $N$ input copies and in the re-preparation of  an $M$-qubit output state $|\Phi_{\hat{\theta}}\>$ conditional  to the estimate  $\hat{\theta}$. In order to maximize the global fidelity, the states $|\Phi_{\hat{\theta}}\> =  U_{\hat \theta}^{\otimes N}  |\Phi\>$  should be contained in the symmetric space spanned by the Dicke states $\{|M/2,m\>~  |~  m  =  -M/2 ,\dots,  M/2 \}$, i.e. $|\Phi_{\hat{\theta}}\>=U_{\hat{\theta}}^{\otimes M}|\Phi\>$ with 
\begin{align}\label{foransatz}
|\Phi\>=\sum_{m=-M/2}^{M/2} \sqrt{p_{M,m}}|M/2, m\>,
\end{align} 
for some suitable coefficients $\{p_{M,m} \}$ that can be chosen to be positive without loss of generality.  For states of this form, the optimal covariant POVM  is known  \cite{Holevo} and is given by  $P_{\hat{\theta}}=U_{\hat{\theta}}^{\otimes N} \eta  U_{\hat{\theta}}^{\dag \otimes N}$  where the seed $\eta$ is the rank-one operator $\eta    =  |\eta\>\<  \eta|$ with 
\begin{align}\label{etaqubit} |\eta\>:=\sum_{n=-N/2}^{N/2}|N/2, n\>.
\end{align} 

Now, the expression for the fidelity is given by  
\begin{equation}\label{eq:m&p-fidelitytheta}
F [  N \to M]=\int \frac{\d \theta}{2\pi} ~    \Tr \left[  \eta~  \psi_\theta^{\otimes N}\right]  ~  \Tr \left[ \Phi~  \psi_\theta^{\otimes M}\right]  .  
\end{equation}
and the goal is to maximize it over all possible choices for the coefficients  in Eq. \eqref{foransatz}. The optimization  can be carried out for given values of $N$ and $M$.  However,  the full optimization is not needed if one just wants to discuss the large $M$ asymptotics.    To this purpose, we make a variational ansatz for the coefficients     $\{p_{M,m}\}$ and later we will prove that asymptotically the ansatz is not too restrictive, because it allows one to achieve the fidelity of the optimal cloner. Our variational ansatz  is the following:
\begin{equation}\label{ansatzqubit}
p_{M,m} (\lambda)=
\begin{cases}
b_{\lceil M/\lambda \rceil,m} & -\frac{  \lceil  M/\lambda  \rceil}{2}\leqslant m\leqslant\frac{\lceil M/\lambda\rceil }{2}\\
0 & \text{otherwise}
\end{cases}
\end{equation}
for some $\lambda\geqslant 1$.  We denote by $|\Phi (\lambda)\>$  the state in Eq. (\ref{foransatz}) with the above choice of coefficients.

With our variational choice, the expression for the fidelity in Eq. (\ref{eq:m&p-fidelitytheta})  can be simplified  in the regime  
\begin{equation}\label{rest-lambda}
\frac {  M} {1+\lambda}\gg N. 
\end{equation}
Indeed, under this condition the function $\Tr \left[  \eta  ~ \psi_\theta^{\otimes N}\right]$ varies slowly with respect to $\Tr\left[ \Phi (\lambda)~  \psi_\theta^{\otimes M}\right]$ (see the proof in the Appendix) and therefore we can approximate Eq. (\ref{eq:m&p-fidelitytheta}) with 
\begin{align}
\nonumber F [  N \to M]  &\approx   \Tr \left[  \eta~  \psi^{\otimes N}   \right]  ~   \left(   \int \frac{\d \theta}{2\pi} ~  \Tr \left[ \Phi (\lambda)~  \psi_\theta^{\otimes M}\right]    \right) \\
\label{last} 
&  =     p^{(N)}_{true}   ~  \<\Phi (\lambda)|     \rho^{(M)}_{aver}  |\Phi(\lambda)\>  
\end{align}
where  $\rho_{aver}^{(M)}=\sum_{m=-M/2}^{M/2}b_{M,m}|M/2, m\>\<M/2, m|$ and $p^{(N)}_{true}=\left(\sum_n\sqrt{b_{N,n}}\right)^2.$

If there were no constraint on $|\Phi (\lambda)\>$,  the optimal choice  that maximizes the expectation value $\< \Phi (\lambda)| \rho_{aver}^{(M)} |\Phi (\lambda)\>$ would be $|\Phi (\lambda)\>  =  |M/2, 0\>$, the eigenvector corresponding to the maximum eigenvalue of $\rho_{aver}^{(M)}$.  However,  from Eq. (\ref{ansatzqubit}) it is clear that this would require $M/\lambda   < 1$, in contradiction with the    condition  $ M/(1+ \lambda) \gg N $, under which Eq. (\ref{last}) was derived.    What can be done instead is to choose $\lambda$ in such a way that both conditions  $\lambda  \gg 1$    and $ M/(1+\lambda) \gg N $ are satisfied.  With this choice, the expectation value  $\< \Phi (\lambda )| \rho_{aver}^{(M)} |\Phi (\lambda)\>$ is still close to the maximum eigenvalue:   
\begin{align*}
\< \Phi (\lambda)| \rho_{aver}^{(M)} |\Phi (\lambda)\>  =      \sum_{m  =  -\lceil M/\lambda  \rceil/2}^{\lceil M/\lambda  \rceil/2} b_{M,m}   b_{\lceil M/\lambda \rceil, m}    \approx  b_{M,0} \qquad \lambda \gg 1 
\end{align*} 
Hence, the fidelity of our variational measure-and-prepare protocol, denoted by $F_{\lambda} [N \to M]$, becomes 
\begin{align}\label{eq:qubit m-p fidelity} F_\lambda [N\to M ]  \approx    \< \Phi (\lambda )| \rho_{aver}^{(M)} |\Phi (\lambda)\>   ~p^{(N)}_{true}\approx b_{M,0}\left(\sum_{n=-N/2}^{N/2}\sqrt{b_{N,n}}\right)^2  \approx F_{clon}  [N\to M],
\end{align}
where the last approximate equality comes from  Eq. \eqref{eq:qubit opt fidelity}.  
Since by definition the maximum  fidelity $F_{est}[N\to M]$ over all measure-and-prepare channels is lower bounded by $  F_\lambda[N \to M]$ and upper bounded  by  $F_{clon}  [N\to M]$, we conclude that 
\begin{align*}
 F_{est}  [N \to M]     \approx  F_\lambda[N\to M]   \approx   F_{clon} [N \to M]  \qquad  \frac M{1+\lambda}  \gg N  ,   \lambda \gg 1.  
 \end{align*}
 This shows that 
 asymptotically, there is no loss of generality in our ansatz: the protocols satisfying the ansatz have a fidelity that is arbitrarily close to the fidelity of the best measure-and-prepare protocol, which in turn is asymptotically equal to the fidelity of the best quantum cloner.       


Let us consider now the fidelity of the naive measure-and-prepare protocol that consists in estimating the phase $\theta$ and re-preparing $M$ identical copies of the estimated state.  In this case, we have  $|  \Phi\>  =  |\psi\>^{\otimes M}   \equiv  |  \Phi(\lambda = 1)\>$
[cf. Eqs. (\ref{foransatz}) and (\ref{ansatzqubit})], and, therefore,  $\< \Phi (\lambda =1)| \rho_{aver}^{(M)} |\Phi (\lambda=1)\>  =      \sum_{m  =  -M/2}^{M/2} b^2_{M,m} $.

For large $M$, the Gaussian approximation gives  $\< \Phi (\lambda =1)| \rho_{aver}^{(M)} |\Phi (\lambda=1)\>   \approx  \sqrt{ 1/(  \pi M)}  \approx  b_{M,0}/\sqrt{2}$, and the fidelity becomes  
\begin{align*}
F_{\lambda =1}  [N \to M]  \approx    \frac{ b_{M,0}}{\sqrt{2}}~\left(\sum_{n=-N/2}^{N/2}\sqrt{b_{N,n}}\right)^2  \approx \frac{F_{clon} [N \to M]}{\sqrt 2} \qquad \forall N \in \mathbb N.
\end{align*}
This proves that re-preparing $M$ identical copies is a strictly suboptimal strategy, which cannot reach the global fidelity of the optimal cloner.

In summary, in this section we showed that the fidelity of the optimal quantum cloner $F_{clon} [N\to M]$ is asymptotically equal to the fidelity of the optimal measure-and-prepare protocol $F_{est} [N\to M]$ in the  limit $M \to \infty$.   Hence, the conjectured equality in Eq. (\ref{conj1}) is verified.   However, achieving the optimal fidelity requires one to prepare suitable $M$-partite entangled states:   the simple strategy consisting in re-preparing $M$ identical copies of estimated state does not give the maximal fidelity,  {\em even in the asymptotic limit}.

\section{Cloning two-qubit maximally entangled states}\label{sec:entangle}

In this section we consider the  N-to-M cloning of two-qubit maximally entangled states, computing the fidelity of the optimal economical covariant cloner and showing that it can be asymptotically attained via a suitable measure-and-prepare protocol.

Consider a general two-qubit maximally entangled state  $|\psi_g\>  \in  \spc H_A \otimes \spc H_B, ~  \spc H_A    \simeq  \spc H_B \simeq  \Cmplx^2$, which can be parametrized as
\begin{equation}
|\psi_g\>=\frac {(U_g\otimes I)|I\>\!\>}{\sqrt{2}}, \qquad g\in SU(2).
\end{equation}
Here we are using the ``double-ket notation"  $  |A\>\!\>   :=   \sum_{m,n}    \< m  |  A    |  n\>   ~  |m\>  |n\>$ for a generic operator $  A  \in  \Lin  (  \spc H)$ \cite{DarianoLoPresti00}.     

We now give a convenient decomposition of the input state  $|\psi_g\>^{\otimes N}  = \left( \spc H_A  \otimes \spc H_B \right)^{\otimes N}  \simeq  \spc H_A^{\otimes N}  \otimes  \spc H_B^{\otimes  N} $.      
With a suitable choice of basis, the Hilbert space $\spc H_A^{\otimes N}$, can be decomposed as  a direct sum of tensor product pairs
\begin{equation}\label{cgh}
\spc H_A^{\otimes N} = \bigoplus_{j   =j_{min}^{(N)}  }^{N/2}    \left(   \spc R^{(j,N)}_{A} \otimes \spc M^{(j,N)}_{A} \right),
\end{equation} 
where    $j$ is the quantum number of the total angular momentum and $j_{min}^{(N)}=0$ for even $N$ while $j_{min}^{(N)}=\frac{1}{2}$ for odd $N$,   $ \spc R^{(j,N)}_{A}$ is a representation space, of dimension  $d_j = 2 j +1$, and $\spc M^{(j,N)}_{A}$ is a multiplicity space, of dimension  $m_j^{(N)}=\frac{2j+1}{N/2+j+1}{N\choose N/2+j}$    (see e.g. Ref. \cite{ChiribellaDariano05}). Relative to this decomposition, we can express $U_g^{\otimes N}$ as a block diagonal matrix, where each  block corresponds to an irreducible representation of $SU(2)$, namely
\begin{equation}\label{cg}
U_g^{\otimes N}=\bigoplus_{j=j_{min}^{(N)}}^{N/2} \left[  U_{g}^{(j,N)}\otimes I_{m_j}^{(N)} \right].
\end{equation}
where $U_g^{(j,N)}  \in\Lin  (\spc R^{(j,N)}_{A}) $ is the unitary operator representing the action of the element $g\in  SU(2)$  and  $I_{m_j}^{(N)}$ denotes the identity on   $\spc M^{(j,N)}_{A}$.

Using  Eq. (\ref{cg}),   the input state   $|\psi_g\>^{\otimes N}$ can be cast in the form
\begin{equation}
\begin{split}
|\psi_g\>^{\otimes N}&=2^{-N/2}(U_g\otimes I)^{\otimes N}|I\>\!\>^{\otimes N} =2^{-N/2}\bigoplus_{j=j_{min}^{(N)}}^{N/2}   \left(|U_{g}^{(j,N)}\>\!\>\otimes|I_{m_j}^{(N)}\>\!\>\right)\\
\end{split}
\end{equation}
with $ |U_{g}^{(j,N)}\>\!\>  \in \spc R^{(j,N)}_{A}  \otimes \spc R^{(j,N)}_{B} $ and $|I_{m_j}^{(N)}\>\!\>  \in \spc M^{(j,N)}_{A} \otimes \spc M^{(j,N)}_{B} $.   Hence, we obtained the decomposition 
\begin{equation}\label{decompositionmaxent} 
|\psi_g\>^{\otimes N}=\bigoplus_{j=j_{min}^{(N)}}^{N/2}\sqrt{c_{j}^{(N)}}|\psi_{g}^{(j,N)}\>\qquad |\psi_{g}^{(j,N)}\>:=\frac{|U_{g}^{(j,N)}\>\!\>}{\sqrt{d_j}}\otimes\frac{|I_{m_j}^{(N)}\>\!\>}{\sqrt{m_j^{(N)}}}
\end{equation}
and $c_{j}^{(N)}:=\frac{d_j m_j^{(N)}}{2^N}=\frac{(2j+1)^2}{(N/2+j+1)}  ~ b_{N,j}$,   
$b_{N,j}$ being the binomial distribution  $b_{N,j}  =  {N \choose  N/2 + j}  /  2^N$.
Note that every state $ |  \psi_g\>^{\otimes N}$ in Eq. (\ref{decompositionmaxent}) belongs to the  subspace 
\begin{align}
 \spc H^{(N)}_{ent}  :  =    \bigoplus_{j={j_{min}^{(N)}}}^{N/2}  \left( \spc R_A^{(j,N)}  \otimes  \spc R_B^{(j,N)}   \otimes  \spc M_A^{(j,N)}  \otimes  \spc M_B^{(j,N)}\right)  \subset (\spc H_A \otimes \spc H_B)^{\otimes N}.
\end{align}   

Hence, for the optimization of the fidelity  we can restrict our attention to this subspace and consider quantum channels that map  states on $\spc H^{(N)}_{ent}$ to states on $\spc H^{(M)}_{ent}$. 

\subsection{The performance of the optimal economical covariant cloner}
Here we focus here on a special type of cloning machines, namely economical covariant cloning machines \cite{NiuGriffiths99,BuscemiDariano05,DurtFiurasek05}.   An economical covariant cloner is described by an isometric channel $  \map C  (\rho)  =   V \rho V^\dag$, where  $V:    \spc H_{ent}^{(N)}\to \spc H_{ent}^{(M)} $ is an isometry satisfying  the covariance requirement, which in our case is expressed by the relation  
 \begin{align}\label{bbb}
  (U_g  \otimes U_h)^{\otimes M}V\left(U_g^{\dag}\otimes U^\dag_h\right)^{\otimes N}=V \qquad \forall g, h\in \grp G.
  \end{align} 
 Note that the action of $ (U_g  \otimes U_h)^{\otimes N}$, restricted to the subspace $\spc H_{ent}^{(N)}$ is  
 \begin{align*}
 \left. (U_g  \otimes U_h)^{\otimes N} \right|_{\spc H_{ent}^{(N)}} =       \bigoplus_{j={j_{min}^{(N)}}}^{N/2}  \left (  U_g^{(j,N)}\otimes U_h^{(j,N)}\otimes I_{m_j,A}^{(N)}\otimes  I_{m_j,B}^{(N)}  \right),
 \end{align*}  
where   $I_{m_j,A}^{(N)}  $   ($ I_{m_j,B}^{(N)}$)  denotes the identity on the multiplicity space $\spc M_A^{(j,N)}$  ($\spc M_B^{(j,N)}$).  
A similar decomposition holds for  the action of $ (U_g  \otimes U_h)^{\otimes M}$ restricted to the subspace $\spc H_{ent}^{(M)}$.  

Now, using the Schur's lemma, Eq. (\ref{bbb}) is reduced to  
\begin{equation}\label{inviso}
V=\bigoplus_{j={j_{min}^{(N)}}}^{N/2}(R_j\otimes M_j),
\end{equation}
where $R_j  :    \spc R_A^{(j,N)}  \otimes  \spc R_B^{(j,N)}  \to \spc R_A^{(j,M)}  \otimes  \spc R_B^{(j,M)} $ is an isometry acting on the representation spaces and satisfying    
$\left( U_g^{(j,M)} \otimes U_h^{(j,M)}  \right) R_j  \left( U_g^{(j,N)} \otimes U_h^{(j,N)}  \right)=  V \quad \forall g, h \in\grp G,$  and  $M_j  :    \spc M_A^{(j,N)}  \otimes  \spc M_B^{(j,N)}  \to \spc M_A^{(j,M)}  \otimes  \spc M_B^{(j,M)} $ is an  isometry acting on the multiplicity spaces. 
 
The fidelity of the economical cloner with isometry $V$ in Eq. (\ref{inviso}) is given by
\begin{align*}
F[N\to M] & =  \left|     \<  \psi |^{\otimes M}   ~ V |  \psi\>^{\otimes N} \right|^2   =  \left|\sum_{j=j_{min}^{(N)}}^{N/2}\sqrt{c_j^{(N)}c_j^{(M)}}\<\psi^{(j,N)}|V|\psi^{(j,M)}\>\right|^2  \le\left(\sum_{j=j_{min}^{(N)}}^{N/2}\sqrt{c_{j}^{(N)}c_{j}^{(M)}}\right)^2.
\end{align*}
The equality holds when   $V  | \psi^{(j,N)}\> = |\psi^{(j,M)}\>$ for every  $j=j_{min}^{(N)}, \dots, {N/2}$, or, equivalently,  
\begin{align*}
  R_j  \frac{|I^{(j,N)}\>\!\>}{\sqrt{d_j}}  =   \frac{|I^{(j,N)}\>\!\>}{\sqrt{d_j}} ,
 \qquad    M_j  \frac{|I_{m_j}^{(N)}\>\!\>}{\sqrt{m_j^{(N)}}}
 = \frac{|I_{m_j}^{(M)}\>\!\>}{\sqrt{m_j^{(M)}}}
 \end{align*}
Interestingly, the optimal covariant economical cloner can be achieved using local operations, because the above isometries only represent an embedding of the state of  $N$ systems on $A$'s and $B$'s sides  into Hilbert space of $M$ systems, and these embedding operations can be carried out locally.    

The maximal fidelity among all possible economical covariant cloner is then
\begin{equation}\label{ccc}
F_{eco,clon}[N \to M]=\left(\sum_{j=j_{min}^{(N)}}^{N/2}\sqrt{c_{j}^{(N)}c_{j}^{(M)}}\right)^2,
\end{equation}
and,  when $M$ is large compared to $N$, becomes
\begin{align}\label{eq:entangle opt fidelity}
   F_{eco,clon}[N\to M]\approx  \frac{2b_{M,0}}{M}     \left(\sum_{j=j_{min}^{(N)}}^{N/2}  \sqrt{\frac{b_{N,j}  (2j+1)^4}{N/2+j+1}}\right)^2  \quad M\gg N.
\end{align}
In the next subsection we will construct a measure-and-prepare channel that achieves this asymptotic value, despite the fact that the fact that the cloning machine considered here is economical, and, therefore, far from  a measure-and-prepare channel.

Before concluding, it is worth noting that the expression for the optimal quantum fidelity becomes simpler when $N$ is large ($N\gg 1$).   Approximating the summation in Eq. \eqref{ccc} with a Gaussian integral,  one obtains  the fidelity 
\begin{equation}
\begin{split}
F_{eco,clon}[N \to M] & \approx\left(\int_{0}^{N/2}8x^2\sqrt{\frac{g_M(x)g_N(x)}{MN}} \d x\right)^2  \approx\left(\frac{4N}{M}\right)^{3/2}\quad N\gg 1.
\end{split}
\end{equation} 
Also in this case, it is also interesting to observe that the fidelity is close to 1 whenever  the number of extra-copies $M-N$ is negligible compared to $N$, whereas it is close to 0 whenever $N$ is negligible compared to $M$, in agreement with the standard quantum limit for cloning  \cite{ChiribellaYang13}.

\subsection{A family of measure-and-prepare protocols achieving asymptotically the optimal fidelity}
Here we  show how to reach the fidelity $F_{clon,eco} [N\to M]$  with a suitable measure-and-prepare  protocol.
Also in this case, we will first make a series of assumptions on the  protocol, and we will eventually show that asymptotically our choice achieves the desired fidelity.

To start with, we consider strategies where the states re-prepared are of the form 
\begin{equation}\label{Phi}
|\Phi_{\hat{g}}\>=\bigoplus_{j=j_{min}^{(M)}}^{M/2}\sqrt{p_{j}^{(M)}}  ~  |  \psi_g^{(j,M)}\>  
\end{equation} 
where $|\psi_g^{(j,M)}\>$ is the vector defined in Eq. (\ref{decompositionmaxent}) and  $\{p_j^{(M)} \}$ are some non-negative coefficients.  Our choice is quite natural,  as it is motivated by the form of the desired states   $|\psi_g\>^{\otimes M}$ [cf. Eq. (\ref{decompositionmaxent}) ].   

Once we assume states of this form for the re-preparation, the optimal POVM  for the measurement is known  \cite{ChiribellaDariano05} and is given by the square-root measurement \cite{HausladenWootters94}, which in this case has the expression $P_{\hat{g}}=U_{\hat{g}}^{\otimes N} \eta  U_{\hat{g}}^{\dag \otimes N}$, where $\eta    =  |\eta\>\<  \eta|$ and $|\eta\>=\bigoplus_{j=j_{min}^{(N)}}^{N/2}  d_j   ~   |  \psi^{(j,N)}\>  $

Then, we make  a variational anstaz on the form of the coefficients $\{p_j^{(M)} \}$ in Eq. (\ref{Phi}),   similar to the ansatz made in subsection \ref{subsec:qubitmeasprep}:  we assume
\begin{equation}\label{p}
p_j^{(M)}  (\lambda)=
\begin{cases}
c_j^{(\lceil  M/\lambda\rceil) } & j\leqslant     \lceil \frac{M}{2\lambda}\rceil \\
0 & j>\lceil \frac{M}{2\lambda}\rceil,
\end{cases}
\end{equation}
for a parameter $\lambda \ge 0$ to be optimized.  Denoting by $|\Phi (\lambda)\>$ the state of Eq. (\ref{Phi}) with the variational choice of coefficients, one can argue that asymptotically $|\<\eta|\psi_g\>^{\otimes N}|^2$ varies slowly with respect to $|\<\Phi(\lambda)|\psi_g\>^{\otimes M}|^2$ provided that  $M/(1+\lambda)\gg N$, following the same lines 
illustrated in the Appendix for the case of equatorial qubits.   Hence,  the fidelity can be turned into Eq. \eqref{eq:m&p-approx} with:
\begin{equation}
\rho_{aver}^{(M)}=\sum_{j=j_{min}^{(M)}}^{M/2}  {c^{(M)}_j}   ~  \left[   \frac {  I_A^{(j,M)}}  {d_j}    \otimes   \frac {  I_{B}^{(j,M)}}  {d_j}  \otimes    \frac{|  I_{m_j}^{(M)} \>\!\>\<  \! \<     I_{m_j}^{(M)} |}{  m_{j}^{(M)}}  \right]
\end{equation}
where   $ I_A^{(j,M)}$   ( $I_B^{(j,M)}$)   denotes the identity on the representation space $\spc R_A^{(j,M)}$  ($\spc R_B^{(j,M)}$),  and  
\begin{equation}
p_{true}^{(N)}=\left(\sum_{j=j_{min}^{(N)}}^{N/2}\sqrt{c_j^{(N)}}d_j\right)^2.
\end{equation}

With similar observation as in subsection \ref{subsec:qubitmeasprep},  $\lambda$ should be chosen in such a way that both conditions  $\lambda  \gg 1$   and $ M/\lambda \gg N $ are satisfied.  With this choice, the expectation value  $\< \Phi (\lambda )| \rho_{aver}^{(M)} |\Phi (\lambda)\>$ is
\begin{align*}
\< \Phi (\lambda)| \rho_{aver}^{(M)} |\Phi (\lambda)\>  &=      \sum_{j  =  j_{min}^{(\lceil M/\lambda  \rceil)}}^{\lceil M/\lambda  \rceil/2} \frac{ c_j^{(\lceil M/\lambda  \rceil)}c_j^{(M)} }{d_j^2}   \approx \frac{2b_{M,0}}{M}   \qquad \lambda \gg 1 
\end{align*} 

Hence, the fidelity of our measure-and-prepare protocol, denoted by $F_{\lambda} [N \to M]$, becomes 
\begin{equation}\label{eq:entangle m-p fidelity}
\begin{split}
F_\lambda [N\to M ]&  \approx    \< \Phi (\lambda )| \rho_{aver}^{(M)} |\Phi (\lambda)\>   ~p^{(N)}_{true} \approx \frac{2b_{M,0}}{M}\left(\sum_{j=j_{min}^{(N)}}^{N/2}(2j+1)^2\sqrt{\frac{b_{N,j}}{N/2+j+1}}\right)^2 \\
&   \approx F_{eco,clon}  [N\to M],
\end{split}
\end{equation}
 the last approximate equality coming from  Eq. \eqref{eq:entangle opt fidelity}.   This shows that asymptotically,  the fidelity of the protocols satisfying the assumptions gets arbitrarily close to the fidelity of the best economical covariant cloner.       

Also in this case,  we can compare the fidelity of our measure-and-prepare protocol with the naive protocol  that consists in estimating the  state and re-preparing $M$ identical copies according to the estimate.  In this case, we have 
$|  \Phi\>  =  |\psi\>^{\otimes M}   \equiv  |  \Phi(\lambda = 1)\>$ [cf. Eqs. (\ref{Phi}) and (\ref{p})], and, therefore, 
\begin{align}
\< \Phi (\lambda =1)| \rho_{aver}^{(M)} |\Phi (\lambda=1)\>  =      \sum_{j=j_{min}^{(M)}}^{M/2} \left(\frac{c_j^{(M)}}{d_j}\right)^2 .
\end{align} 
For large $M$, this gives  $\< \Phi (\lambda =1)| \rho_{aver}^{(M)} |\Phi (\lambda=1)\>   \approx  \sqrt{ 1/(  \pi M^3)}  \approx  b_{M,0}/(\sqrt{2}M)$, and the fidelity becomes  
\begin{align*}
F_{\lambda =1}  [N \to M]  \approx    \frac{ b_{M,0}}{\sqrt{2}M}~\left(\sum_{j=j_{min}^{(N)}}^{N/2}\sqrt{c_j^{(N)}}d_j\right)^2  \approx \frac{F_{eco,clon} [N \to M]}{2^{3/2}}.
\end{align*}
This proves that re-preparing $M$ identical copies is a strictly suboptimal strategy, which cannot reach the  fidelity of the optimal economical covariant cloner.

In summary, in this section we showed that for the case of two-qubit maximally entangled states, the fidelity of the optimal economical covariant cloner $F_{eco,clon} [N\to M]$  can be achieved   by a measure-and-prepare protocol $F_{\lambda} [N\to M]$ in the asymptotic limit $M \to \infty$.     However, achieving the optimal fidelity requires one to prepare suitable $M$-partite entangled states:   the simple strategy consisting in re-preparing $M$ identical copies of estimated state does not give the desired fidelity, even in the asymptotic limit.

\section{Discussion and conclusions}\label{sec:conclusion}

In this paper we posed the question whether the asymptotic cloning is equivalent to state estimation in terms of the global fidelity between the output state of all clones and the desired state of $M$ perfect copies.  
To gain insight into the problem, we provided two examples (cloning of equatorial qubit states and cloning of two-qubit maximally entangled states) where the equivalence between cloning and estimation is satisfied in a rather non-trivial way, despite the cloning machines under consideration are economical.    Our results suggest the existence of a general mechanism that guarantees the equality of fidelities in Eq. (\ref{conj1}).  Finding a general proof, or finding a counterexample to the conjectured equivalence between global asymptotic cloning and state estimation is the most pressing open question raised by our work.

\subparagraph*{Acknowledgements}  This work is supported by the National Basic Research Program of China (973) 2011CBA00300 (2011CBA00301), by the National Natural Science Foundation of China through Grants 61033001,  61061130540, and 11350110207, and by the 1000 Youth Fellowship Program of China.

\bibliography{ref}

\appendix

\section{Justification of the asymptotic approximation for the fidelity}
In order to study the asymptotic behavior of the fidelity for measure-and-prepare protocols, we can use a Taylor expansion of $\Tr \left[  \eta~  \psi_g^{\otimes N}\right]$ up to the second order term:
\begin{equation}
\begin{split}
\Tr \left[  \eta~  \psi_g^{\otimes N}\right]&=\sum_{n,m=-N/2}^{N/2}\sqrt{b_{N,n}b_{M,m}}e^{i(n-m)\theta}\\
& \approx \left(\sum_{n=-N/2}^{N/2}\sqrt{b_{N,n}}\right)^2-\frac{\theta^2}{2}\left[\sum_{n,m=-N/2}^{N/2}\sqrt{b_{N,n}b_{N,m}}(n-m)^2\right]\\
&= \left(\sum_{n=-N/2}^{N/2}\sqrt{b_{N,n}}\right)^2-\theta^2\left(\sum_{n=-N/2}^{N/2}n^2\sqrt{b_{N,n}}\right)\left(\sum_{m=-N/2}^{N/2}\sqrt{b_{N,m}}\right).
\end{split}
\end{equation}
In the asymptotic limit of large amplification, i.e. $M\gg N$, $b_{M,m}$ can be approximated by the Gaussian $g_M(x):=\sqrt{\frac{2}{\pi M}}e^{-\frac{2x^2}{M}}$, thus we have:
\begin{align*}
\Tr \left[ \Phi~  \psi_g^{\otimes M}\right]&= \left|\sum_{m=-\lceil M/2\lambda \rceil}^{\lceil M/2\lambda \rceil}\sqrt{b_{\lceil M/\lambda \rceil,m}b_{M,m}}e^{im\theta}\right|^2 \approx \left|\int_{-\lceil M/2\lambda \rceil}^{\lceil M/2\lambda \rceil}\d x~\sqrt{\frac{2}{\pi M}}~\lambda^{\frac{1}{4}}~e^{-\frac{(1+\lambda)x^2}{M}+i\theta x}\right|^2\\
&=\frac{2\sqrt{\lambda}}{1+\lambda}~e^{-\frac{M\theta^2}{2(1+\lambda)}}.
\end{align*}
Taking these into Eq. \eqref{eq:m&p-fidelity} we get the expression of measure-and-prepare fidelity for large amplification:
\begin{align}
\nonumber F[N\to M]&=\int \frac{\d \theta}{2\pi} ~    \Tr \left[  \eta~  \psi_\theta^{\otimes N}\right]  ~  \Tr \left[ \Phi~  \psi_\theta^{\otimes M}\right]\\
\nonumber &\approx \frac{2\sqrt{\lambda}}{1+\lambda}~\left(\sum_{n=-N/2}^{N/2}\sqrt{b_{N,n}}\right)\left(\sum_{n=-N/2}^{N/2}\sqrt{b_{N,n}}~\int_{-\pi}^{\pi}\frac{\d \theta}{2\pi}~e^{-\frac{M\theta^2}{2(1+\lambda)}}\right.\\
\nonumber &\left.-\sum_{n=-N/2}^{N/2}n^2\sqrt{b_{N,n}}~\int_{-\pi}^{\pi}\frac{\d \theta}{2\pi}~\theta^2~e^{-\frac{M\theta^2}{2(1+\lambda)}}\right)\\
&=\sqrt{\frac{2\lambda}{\pi M(1+\lambda)}}~\left(\sum_{n=-N/2}^{N/2}\sqrt{b_{N,n}}\right)\left(\sum_{n=-N/2}^{N/2}\sqrt{b_{N,n}}-\frac{1+\lambda}{2\pi M}\sum_{n=-N/2}^{N/2}n^2\sqrt{b_{N,n}}\right).
\label{eq:append-m&p-fidelity}
\end{align}
It is clear that the contribution resulting from the second order term of the expansion is negligible whenever $M/(1+\lambda)  \gg N$.  
In particular, for large $N$  the binomial can be approximated by a Gaussian, 
giving
\begin{align}
\sum_{n=  - N/2}^{N/2}  \sqrt{b_{N,n}}     \approx  (2\pi  N)^{1/4}    \qquad 
\sum_{n=  - N/2}^{N/2}  n^2~ \sqrt{b_{N,n}}     \approx ( 2\pi  N)^{1/4}   \frac N 2  .
\end{align}
so that the ratio between the second and first order term in Eq. (\ref{eq:append-m&p-fidelity}) is $N(1+\lambda)/M$.   

\end{document}